\documentclass[12pt,indent]{article}
\usepackage{a4,latexsym,amsmath,amssymb}
\usepackage[latin1]{inputenc}
\usepackage{float}
\usepackage{natbib}
\usepackage{graphics}
\usepackage[english]{babel}
\usepackage{tabularx}
\usepackage{lscape}
\usepackage{multirow}
\usepackage{rotating}
\usepackage{dsfont}
\usepackage{caption}
\usepackage{subcaption}
\usepackage{bbm}
\usepackage[table]{xcolor}
\usepackage{booktabs}
\usepackage{calc}
\usepackage{ifthen}
\usepackage{url}
\usepackage{colortbl}      
\usepackage{tikz}

\usepackage{blindtext}
\usepackage{scrextend}
\usepackage{mathtools}
\usepackage{verbatim}

\usetikzlibrary{shapes,snakes,matrix}

\newenvironment{tabularsmall}
{ \footnotesize \sffamily \tabular } {
\endtabular
\normalfont }

\newcommand{\E}{\operatorname{E}}      
\newcommand{\var}{\operatorname{var}}

\newcommand{\betab}{{\boldsymbol{\beta}}}

\newcommand{\deltab}{\boldsymbol{\delta}}

\newcommand{\gammab}{\boldsymbol{\gamma}}

\newcommand{\lambdab}{\boldsymbol{\lambda}}

\newcommand{\varepsilonb}{{\boldsymbol{\varepsilon}}}

\newcommand{\Phib}{{\boldsymbol{\Phi}}}

\newcommand{\Sigmab}{\boldsymbol{\mathit\Sigma}}

\newcommand{\bb}{\boldsymbol{b}}

\newcommand{\Ib}{\boldsymbol{I}}

\newcommand{\Xb}{\boldsymbol{X}}
\newcommand{\xb}{\boldsymbol{x}}

\newcommand{\yb}{\boldsymbol{y}}

\newcommand{\Zb}{\boldsymbol{Z}}
\newcommand{\zb}{\boldsymbol{z}}

\newcommand{\blanco}[1]{}

\def\d{\displaystyle}

\usepackage{natbib}

\usepackage{geometry}
\usepackage{setspace}
\usepackage{tikz}
\usepackage[]{graphicx}
\usepackage[]{color}

\definecolor{fgcolor}{rgb}{0.345, 0.345, 0.345}

\usepackage{framed}
\makeatletter
 {\par\unskip\endMakeFramed%
 \at@end@of@kframe}
\makeatother

\definecolor{shadecolor}{rgb}{.97, .97, .97}
\definecolor{messagecolor}{rgb}{0, 0, 0}
\definecolor{warningcolor}{rgb}{1, 0, 1}
\definecolor{errorcolor}{rgb}{1, 0, 0}

\usepackage{alltt}
\IfFileExists{upquote.sty}{\usepackage{upquote}}{}

\newtheorem{theorem}{Proposition}[section]

\begin{document}
\bibliographystyle{chicago}
\sloppy

\makeatletter
\renewcommand{\section}{\@startsection{section}{1}{\z@}%
        {-3.5ex \@plus -1ex \@minus -.2ex}%
        {1.5ex \@plus.2ex}%
        {\reset@font\Large\sffamily}}
\renewcommand{\subsection}{\@startsection{subsection}{1}{\z@}%
        {-3.25ex \@plus -1ex \@minus -.2ex}%
        {1.1ex \@plus.2ex}%
        {\reset@font\large\sffamily\flushleft}}
\renewcommand{\subsubsection}{\@startsection{subsubsection}{1}{\z@}%
        {-3.25ex \@plus -1ex \@minus -.2ex}%
        {1.1ex \@plus.2ex}%
        {\reset@font\normalsize\sffamily\flushleft}}
\makeatother



\newsavebox{\tempbox}
\newlength{\linelength}
\setlength{\linelength}{\linewidth-10mm} \makeatletter
\renewcommand{\@makecaption}[2]
{
  \renewcommand{\baselinestretch}{1.1} \normalsize\small
  \vspace{5mm}
  \sbox{\tempbox}{#1: #2}
  \ifthenelse{\lengthtest{\wd\tempbox>\linelength}}
  {\noindent\hspace*{4mm}\parbox{\linewidth-10mm}{\sc#1: \sl#2\par}}
  {\begin{center}\sc#1: \sl#2\par\end{center}}
}



\def\R{\mathchoice{ \hbox{${\rm I}\!{\rm R}$} }
                   { \hbox{${\rm I}\!{\rm R}$} }
                   { \hbox{$ \scriptstyle  {\rm I}\!{\rm R}$} }
                   { \hbox{$ \scriptscriptstyle  {\rm I}\!{\rm R}$} }  }

\def\N{\mathchoice{ \hbox{${\rm I}\!{\rm N}$} }
                   { \hbox{${\rm I}\!{\rm N}$} }
                   { \hbox{$ \scriptstyle  {\rm I}\!{\rm N}$} }
                   { \hbox{$ \scriptscriptstyle  {\rm I}\!{\rm N}$} }  }

\def\d{\displaystyle}\def\d{\displaystyle}

\title{A General Framework for Random Effects Models for Binary, Ordinal, Count Type and Continuous Dependent Variables Including Variable Selection}
\author{Gerhard Tutz et al\\{\small Ludwig-Maximilians-Universit\"{a}t M\"{u}nchen}\\
{\small Akademiestra{\ss}e 1, 80799 M\"{u}nchen}}

\maketitle

\begin{abstract} 
\noindent

A general random effects model is proposed that allows for continuous as well as discrete distributions of the responses. Responses can be unrestricted continuous, bounded continuous, binary, ordered categorical
or given in the form of counts. The distribution of the responses is not restricted to exponential families, which is a severe restriction in generalized mixed models. Generalized mixed models use fixed distributions for responses, for example the Poisson distribution in count data, which has the disadvantage of not accounting for overdispersion. By using a response function and a thresholds function the proposed mixed thresholds model can account for a variety of alternative distributions that often show better fits than fixed distributions used within the generalized linear model framework. A particular strength 
of the model is that it provides a tool for joint modeling, responses may be of different types, some can be discrete, others continuous. 
In addition to introducing the mixed thresholds model parameter sparsity is addressed. 
Random effects models can contain a large number of parameters, in particular if effects have to be assumed as measurement-specific. Methods to obtain sparser representations are proposed and illustrated. The methods are shown to work in  the thresholds model but could also be adapted to other modeling approaches. 

\end{abstract}

\section{Introduction}
Random effects models are a strong tool to model
 the heterogeneity of  clustered responses. By
postulating the existence of unobserved latent variables, the
so-called random effects,  which are shared
by the measurement within a cluster, correlation
between the measurements within clusters is introduced.
The clusters or units can refer to persons in repeated measurement trials or to larger units as, for example, schools with measurements referring to performance scores of students.

Detailed
expositions of linear mixed models, which are typically used for continuous dependent variables, are found in
\citet{Hsiao:86}, \citet{Lindsey:93},  and  \citet{Jones:93}. 
Models for binary variables and counts are often discusses within the framework of generalized mixed models, see, for example, \citet{MccSea:2001}.
Random effects models for ordinal dependent variables  were considered by
\citet{HarMee:84}, \citet{Jansen:90}, \citet{TutHen:96} and \citet{HarLiuAgr:2001}.
Mixed model versions for continuous bounded data in the form of rates and proportions that take values in the interval $(0,1)$, have been considered by  \citet{qiu2008simplex} based on the simplex model and by    \citet{bonat2015likelihood} who propagate beta distribution models. Several R packages are available to fit generalized mixed linaer models, for example, glmmTMB  \citep{brooks2017glmmtmb} for various continuous and discrete  distributions,  lme4  for continuous and binary data, \textit{ordinal} and \textit{MultOrdRS} \citep{schauberger2024package} for ordinal data.




Generalized mixed models within the generalized linear model framework as well as extended approaches as the models for bounded continuous data  postulate  familiar fixed distributions for the responses. This is different in the approach propagated here although there is some overlap with generalized linear models.
The mixed thresholds model used here gains its flexibility concerning distributional assumptions by using two components, a response function, which is a distribution function, and a thresholds function  that modifies the distribution. In the simplest case, by assuming a linear thresholds function,  the distribution of the responses follows the response function. Thus, familiar linear Gaussian response models are obtained but also linear models with quite different, possibly skewed distribution functions are available. Distributions with a restricted support, for example if responses are observed in  an interval or are positive only, are obtained by using non-linear thresholds functions. They are also useful when modeling discrete data, which typically are restricted to a specific range, for example, count data take only  values $0, 1, \dots$ and ordered categorical responses can be coded by  $0, 1, \dots, k$, where numbers only represent the order of values. In the case of binary and ordered categorical responses cumulative generalized mixed model is a special case of the proposed thresholds model.

More concrete, let $y_{i1,\dots,}y_{im}$ denote the observations on unit $i$ ($i=1,\dots,n$),  which can be continuous or discrete. In addition, let $\xb_{ij},\zb_{ij}$ denote  covariates associated with response $y_{ij}$. 
Then, the Mixed Thresholds Model (MTM) has the form

\begin{align}\label{eq:dif}
P(Y_{ij} > y|\bb_i,\zb_{ij},\xb_{ij})=F(\zb_{ij}^T\bb_i+\xb_{ij}^T\betab_j-\delta_{j}(y))),
\end{align}
where $F(.)$ is a strictly increasing distribution function and $\delta_{j}(.))$ is a non-decreasing measurement-specific function defined on the support $S$ of the dependent variables, 
 referred to as \textit{thresholds function}.
The predictor $\eta_{ij}=\zb_{ij}^T\bb_i+\xb_{ij}^T\betab_j-\delta_{j}(y))$
contains  two components   linked to explanatory variables. 

\begin{itemize}
 \item[] The term  $\zb_{ij}^T\bb_i$ contains the cluster-specific effects $\bb_i$. They are assumed to vary independently across clusters and are assumed to follow a specific distribution, typically the normal distribution, $\boldsymbol{b}_i\thicksim N(\boldsymbol{0},\boldsymbol{\Sigma})$.

\item[] The term  $\xb_{ij}^T\betab_j$ contains the effects of $\xb_{ij}$ on the dependent variable. The parameters  $\betab_j$ are fixed population-specific parameters.

\end{itemize}

The distribution of the dependent variables is crucially determined by the choice of the distribution function $F(.)$, also referred to as response function, and the thresholds function $\delta_{j}(.)$. Specific choices yield models that are in common use in random effects modeling. Other choices  widen the toolbox yielding models that show better fits than classical approaches. Thresholds models have been considered before in the form of item thresholds model  \citep{TuItThr2022}, which are latent trait models that aim at measurement and do not contain any covariates. The objective of the models considered here is quite different. The focus is on the effect of covariates on responses, with covariates that can take quite dfferent forms. They can be measurement-specific, unit-specific or both, which raises problems, in particular with the possible number of parameters involved.

The role the response   and the thresholds function play in modeling the response distribution will become obvious when considering specific choices. In Section \ref{sec:cont} the case of continuous responses is considered  with linear and non-linear thresholds functions. Section \ref{sec:discr} is devoted to discrete data with infinite and finite support.
Joint modeling, which allows for different types of responses, in particular a mixture of continuous and discrete responses, are considered in Section  \ref{sec:joint}. Marginal likelihood estimation methods are given in Section \ref{sec:est}. In Section \ref{sec:sparse}
the problem of obtaining sparser representations is addressed.  Several small examples are used   to demonstrate the versatility of the approach. They are   meant for illustration, no in-depth investigation of effects is given.

\section{Continuous Dependent Variables}\label{sec:cont}
 We start with models that contain linear thresholds functions. The  model class comprises the classical normal response model but allows for alternative distributions. Then we consider models for restricted support, which call for non-linear thresholds functions.

\subsection{Linear Random Effects Models for Gaussian Data and Other Distributions}

Let the dependent  variables be continuous with  support $\mathbb{R}$. Then a thresholds function that is simple but already yields  very flexible models is the linear thresholds function 
\[\delta_{j}(y)= \delta_{0j}+ \delta_j y,\quad \delta_j >0. \]
Let $F(.)$ denote a fixed, typically standardized,  distribution function with support $\mathbb{R}$, for example the standardized normal  distribution function.  
Then,   the means $\mu_{ij}=\E(Y_{ij})$  and variances $\sigma_{ij}^2=\var(Y_{ij})$  of dependent  variables have a very simple form,

\begin{align} \label{condex}
&\mu_{ij}=  \frac{1}{\delta_{j}}(\beta_{0j}+\zb_{ij}^T\bb_i+\xb_{ij}^T\betab_j), 
\quad\sigma_{ij}^2=  \frac{\sigma_F^2}{\delta_j^2},
\end{align}
where $\beta_{0j}=-\mu_F-\delta_{0j}$, and $\mu_F,\sigma_F^2$  are constants that are determined by the distribution function $F(.)$. More concrete,  $\mu_F=\int y f(y)dy$ is the expectation corresponding to distribution function $F(.)$ and  $\sigma_F^2=\var_F =\int (y-\mu_F)^2f(y)d y$ the corresponding variance; for a proof see Proposition \ref{t1}.

It is seen that the means of dependent variables are simple linear functions of 
$\zb_{ij},\xb_{ij}$ and the variances vary across measurements. It is a linear model but not necessarily for Gaussian data. Responses can take any strictly increasing distribution function. The model parameterizes the mean as a linear function of covariates also if responses follow a skewed distribution, which might be more appropriate in applications.

The  distribution of responses is easily derived since the density $f_{ij}(.)$ of   variable $Y_{ij}$ is given by
\begin{align} \label{dens}
f_{ij}(y)=f(\zb_{ij}^T\bb_i+\xb_{ij}^T\betab_j-\delta_{0j}- \delta_j y)\delta_j, 
\end{align}
where $f(.)$ denotes the density linked to $F(.)$, $f(y)=\partial F(y)/{\partial y}$. It means, in particular, that for \textit{symmetric} distribution functions $F(.)$ the distributions of all dependent variables are scaled and shifted versions of the distribution specified by $F(.)$. If $F(.)$ is not symmetric the distributions of all dependent variables are scaled and shifted versions of the distribution function $\tilde F(y)=1-F(-y)$. This  is easily seen by considering the distribution function of $Y_{ij} $, which is given by $P(Y_{ij} \le y|\bb_i,\zb_{ij},\xb_{ij})=1-P(Y_{ij} > y|\bb_i,\zb_{ij},\xb_{ij})$.

A simplified version is the \textit{homogeneous} GMT model, in which $\delta_j$ does not depend on the measurement, that is $\delta_1=\dots=\delta_m=\delta$. Then, the variance is the same for all observations  and the mean and variance have the simple form
\begin{align} \label{qu:linbas}
&\mu_{ij}= \tilde\beta_{0j}+\zb_{ij}^T\tilde\bb_i+\xb_{ij}^T\tilde\betab_j,  
\quad \sigma_{ij}^2=   {\tilde\sigma_F^2}, 
\end{align}
where $\tilde\beta_{0j}, \tilde\bb_i,, \tilde\betab_j,\tilde\sigma_F^2$ are the original parameters divided by $\delta$.

A special case of the GMTM is the classical linear random effects model with normally distributed dependent variables. Let $F(.)$ denote the standardized normal distribution function and assume that dispersion homogeneity holds ($\delta_1=\dots=\delta_m=\delta$). Then, the dependent variables are normally distributed with means and variance given by (\ref{qu:linbas}). A more familiar representation of the model is the vector-valued representation
\[
\yb_i=\betab_0+\Xb_i\betab+\Zb_i\bb_i+\varepsilonb_i, \bb_i\sim
N(\boldsymbol{0}, \Sigmab),\varepsilonb_i\sim N(\boldsymbol{0},
\tilde\sigma_F^2\Ib)
\]
where $\boldsymbol{y}_i^T=(y_{i1},\dots,y_{im})$, the  matrices  $\Xb_i,\Zb_i$ are composed from the  vectors $\zb_{ij},\xb_{ij}$ and $\Ib$ denotes the unit matrix.

The more general model with varying dispersion parameters $\delta_j$ is an extension of this classical model. It is more flexible and can be more appropriate, in particular when repeated measurements on a unit are time-dependent and the variance changes over time.

 Within the MTM framework there is no need to assume that dependent variable are normally distributed. Any strictly monotone distribution function $F(.)$ can be used in the model. With linear 
 thresholds function $\delta_{j}(.)$ one obtains a linear form of the expectation and simple terms for the variance. In particular skewed distribution  can be used.
 This extends the usual normal distribution approach to modeling clustered data to a wider class of models with a simple link between covariates and measurements.
 
 \subsubsection*{Rent data}
For illustration we use the Munich rent index data. The variables are 
rent (monthly rent in Euros), 
floor (floor space),
rooms (number of rooms), age in years.
There are 25 districts (residential areas) which have an effect on rents and are modeled as random effects. For an extensive description of the rent data see \citet{fahkne11}.
The rent is the dependent variable, floor space, number of rooms and age are explanatory variables.

Monthly rents  tend to have a right-skewed distribution since there are typically some houses that are much more expensive than the average house. Thus, the normal distribution might not be the best choice for modeling this kind of data. 
A candidate for a right  right-skewed distribution is the Gumbel distribution $F(y)=\exp(-\exp(-y))$, which is the distribution of responses if one chooses 
the Gompertz distribution $F(y)=1-\exp(-\exp(y))$ as response function $F(.)$. 
As the log-likelihoods in Table \ref{tab:rent} show the assumption of the Gumbel distribution for responses (F(.) is chosen as the Gompertz distribution) yields better fit. The Gompertz distribution as a left-skewed distribution for responses (if $F(.)$ is chosen as the Gumbel distribution function ) yields much worse fit and is not shown.

\begin{table}[h!]
 \caption{ Parameter estimates for  self esteem dat } \label{tab:rent}
\centering 
\begin{tabularsmall}{llccccccccc}
  \toprule
 & response distribution   & Floor & Rooms &  Age &std mixture &log-lik \\ 
\midrule
&Normal    &0.073 & -0.593 & -0.010  &0.244  &-3389.992\\
&stderr& ( 0.004)&(0.087) &(0.002)\\
\midrule
&Gumbel    &0.050 &-0.280  & -0.012   &0.331 &-3366.059\\
&stderr& ( 0.004)&(0.099) &(0.002)\\
\bottomrule
\end{tabularsmall}
\end{table}

\subsection{Random Effects Models for Positive-Valued  Variables} 

In many applications the dependent variable can take only positive values, for example if responses are response times. Although often used, the assumption of a normal distribution or any other distribution with support $\mathbb{R}$ is not warranted and will only yield a crude approximation to the true distribution. 

In the MTM the support of the dependent variable can be restricted by using an appropriate thresholds function $\delta_{j}(.)$.  If the difficulty function is chosen such that $\lim_{y\rightarrow 0}\delta_i(y)=-\infty$ holds the responses automatically has positive values, $y \ge 0 $. One candidate that can be chosen is the logarithmic thresholds function 
\[
\delta_{j}(y)= \delta_{0j}+ \delta_j \log(y).
\]
Thresholds functions of this type combine linearity with a transformation function. The general form of thresholds functions of this type, which are used throughout the paper,  is
 \begin{align} \label{genf}
\delta_{j}(y)= \delta_{0j}+ \delta_j g(y),
\end{align} 
where $g(.)$ is a non-decreasing function. Thresholds functions, of this form  are simply named after the transformation function $g(.)$.

If the thresholds function is logarithmic a familiar distribution is found if  $F(.)$ is chosen as the standard normal distribution function. Then, one can derive that the density of $y_{ij}$ denoted by $f_{ij}(.)$ is given by 
\begin{align} 
f_{ij}(y)=
&\frac{1}{\bar\sigma_j y\sqrt{2\pi}} \exp(\frac{-(\log(y)-\tilde\mu_{ij})^2}{ \bar\sigma_j^2})
\end{align} 
where 
$\tilde\mu_{ij}=\zb_{ij}^T\bb_i+\xb_{ij}^T\betab_j-\delta_{0j}\tilde\eta_{ij}\delta_j$, $\bar\sigma_j=1/\delta_j$.
This is the lognormal distribution with parameters 
$\bar\mu_{pi}, \bar\sigma_i$. Thus the logarithmic thresholds function generates a random effects model in which dependent variables follow a lognormal distribution.

\subsubsection*{Sleep Data}
In a sleep deprivation study the average reaction time per day for subjects has been measured (data set sleepstudy from package lme4). On day 1 the subjects had their normal amount of sleep. Starting that night they were restricted to 3 hours of sleep per night. The observations represent the average reaction time on a series of tests given each day to each subject. The 10 days represent the repeated measurements on 18 persons. 

\begin{figure}[h!]
\centering
\includegraphics[width=7cm]{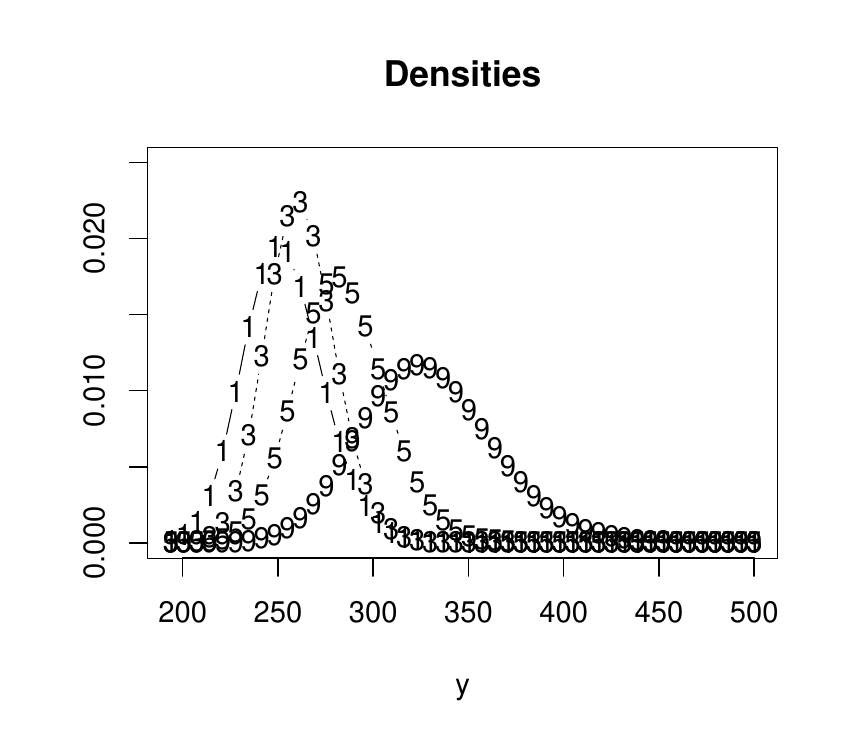}
\includegraphics[width=7cm]{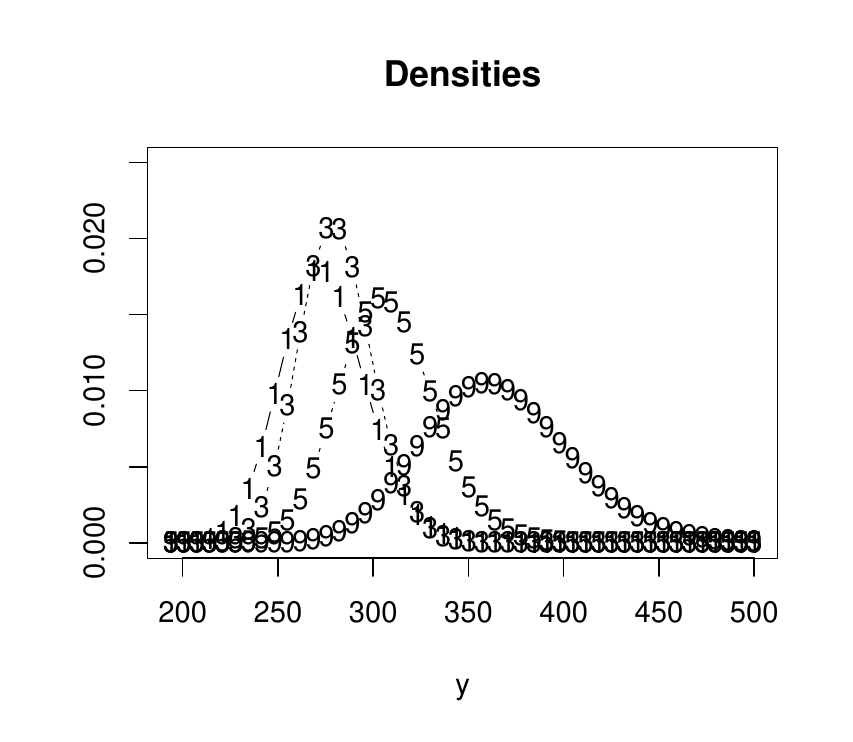}
\caption{Reaction time for days 1,3,5,9 of sleep deprivation for $b_i=0$ (left) and $b_i=1$ (right).}
\label{fig:sleep1}
\end{figure}

Instead of using a normal distribution model with linear effect of days a thresholds model 
 with normal response function and logarithmic thresholds function is used.
 In the model
 \begin{align*}
P(Y_{ij} > y|b_i,\zb_{ij},\xb_{ij})=F(b_i-\delta_{0j}- \delta_j \log(y)),
\end{align*}
the random effect $b_i$ refers to the person and $\delta_{0j}$ accounts for the effect of days. It is not assumed that the mean is a linear function of days as is common in typical random effects models.
Instead, the basic variation of responses over repeated measurements is captured in the parameters $\delta_{0j}$.
The parameters $\delta_j$ account for possible heterogeneity of variances.
The $\log(y)$ function, which makes the dependent variables follow a log-normal distribution shows slightly better fit than the common normal distribution model. The 
 log-likelihood was -872.96 with the $\log(y)$ function and -875.50 for the identity function (Gaussian distribution).   Figure 
\ref{fig:sleep1} shows the fitted densities for  days 1,3,5,9 for $b_i=0$ and $b_i=1$.
It is seen that the distributions have quite different forms and variances vary across days. 
The mean reaction time increases as well as the variances increase over days of sleep deprivation .

\subsection{Random Effects Models for Continuous Bounded Data} 

Various regression model have been proposed for continuous bounded data in the form of rates and proportions that take values in the interval $(0,1)$, see, for example,  \citet{kieschnick2003regression}, \citet{bonat2019flexible}. Also mixed model versions for repeated measurement have been developed, in particular the simplex mixed
model \citep{qiu2008simplex} and   beta mixed models  \citep{bonat2015likelihood}.

Let us more generally consider the case where $Y_{ij}  \in (a,b)$. The restriction of responses to the interval is obtained within the  thresholds model framework by choosing thresholds functions
 for which $\lim_{y\rightarrow a}\delta_i(y)=-\infty$ and $\lim_{y\rightarrow b}\delta_i(y)=\infty$  hold since then $P(Y_{ij} > a)\rightarrow 1$ and $P(Y_{ij} > b)\rightarrow 0$. A thresholds function that meets these demands   is, for example, the logit thresholds function 
 \[
\delta_{j}(y)= \delta_{0j}+ \delta_j \log((y-a)/(b-y)).
\]
Instead of $g(.)=\log((y-a)/(b-y)$  any inverse distribution function can be used. The logistic distribution function is just one option, which yields the logit function  with $g(y)=\log((y-a)/(b-y))$.

While simple terms for means and variances of the dependent variable can be found only for simple thresholds functions as the linear one it is straightforward to show that 
for any (non-decreasing) transformation function $g(y)$ means and expectations of the transformed variables $g(Y_{ij})$ are given by

\begin{align} \label{trans}
&\E(g(Y_{ij}))=  \frac{1}{\delta_{j}}(\beta_{0j}+\zb_{ij}^T\bb_i+\xb_{ij}^T\betab_j), 
\quad\var(g(Y_{ij}))=  \frac{\sigma_F^2}{\delta_j^2},
\end{align}
see Proposition \ref{t2} for a proof. That means the mean of the responses is a linear function of covariates and variances can vary over measurements.

\begin{figure}[h!]
\centering
\includegraphics[width=6cm]{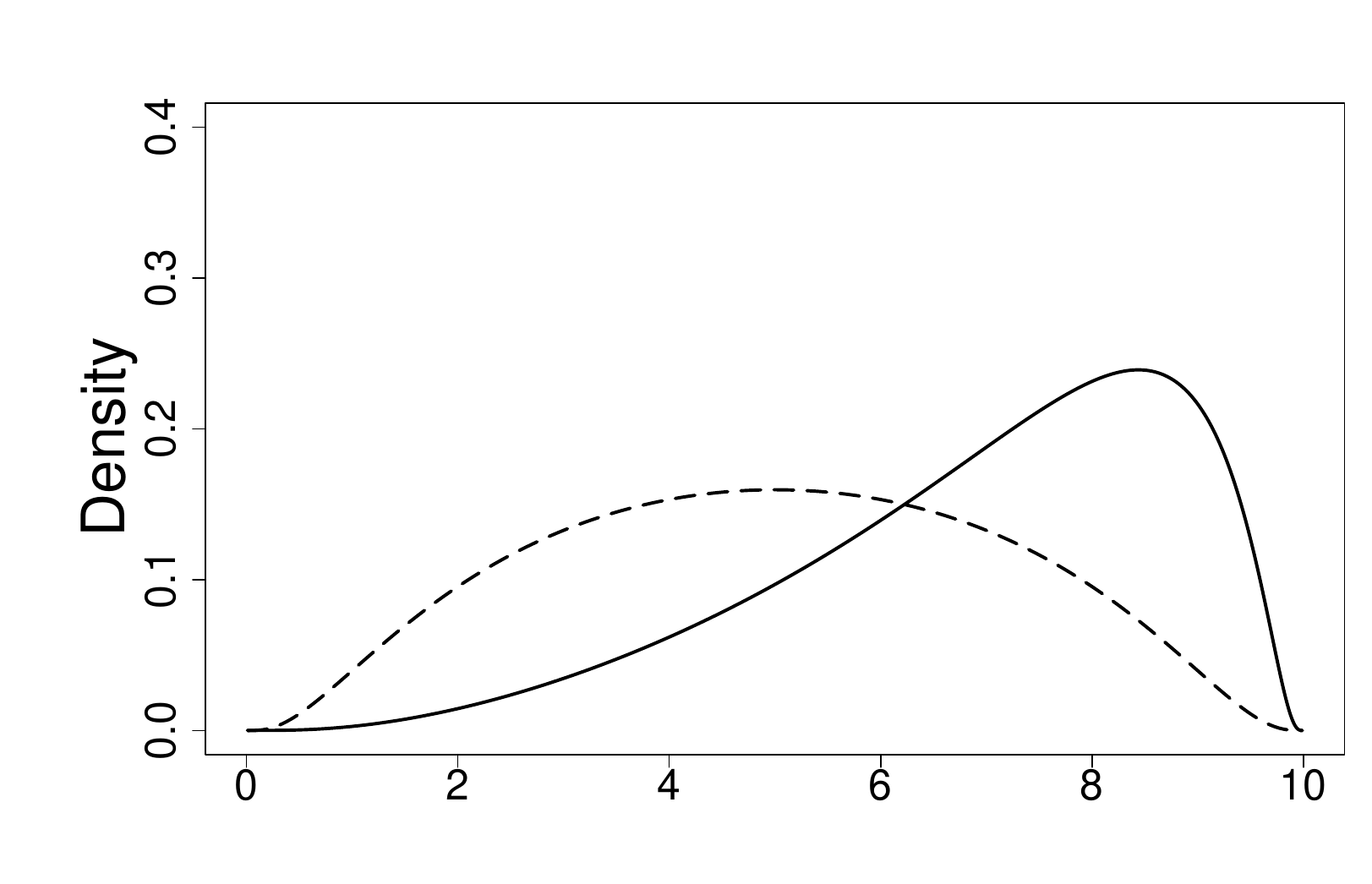}
\includegraphics[width=6cm]{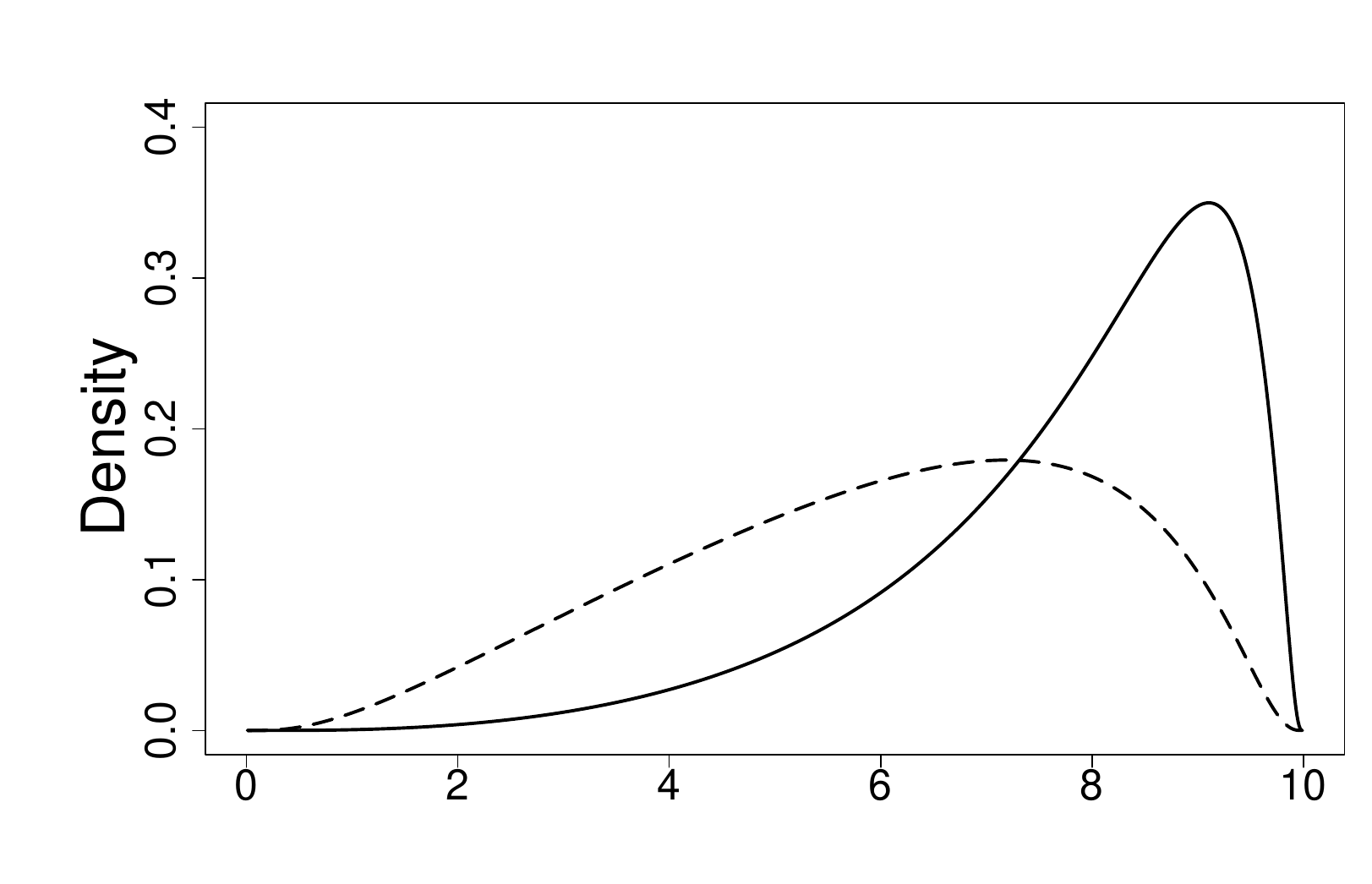}
\caption{Densities for predictor is $\eta_{ij}=b_i+x_i\beta$ with binary predictor $x_i$, $\beta=1$ and $\delta_{j}(y)= \log((y-a)/(b-y)), a=0, b=10$;  left: $b_i=0$;  right: $b_i=0.5$;
drawn line: $x_i=0$,  dashed line: $x_i=1$.}
\label{fig:bound1}
\end{figure}

To illustrate the restriction to intervals generated by properly chosen thresholds functions Figure \ref {fig:bound1} shows the obtained distributions if $F(.)$ is the normal distribution, the predictor is $\eta_{ij}=b_i+x_i\beta$ with binary predictor $x_i$, $\beta=1$ and $\delta_{j}(y)= \log((y-a)/(b-y)), a=0, b=10$. In the left picture  $b_i=0$, in the right picture $b_i=0.5$. The  drawn line shows the density for $x_i=0$, the dashed line  for
$x_i=1$. It is seen that the logit type distribution ensures that the support of the distribution 
is $(0,1)$. For larger random effect (right picture) the density becomes larger close to the upper boundary.

\section{Discrete Data}\label{sec:discr}

Discrete data come in two forms, with infinite support and finite support. We will consider first the case of infinite support than the case where the response is in categories.  In the latter typically only ordinal scale level is assumed for the dependent variable.

\subsection{Random Effects Models for Count Data} 

Let the responses be counts, that is, $Y_{ij}$ takes values from $\{  0,1, \dots\}$. 
If responses are assumed to follow a Poisson distribution mixed models can be formulated within the generalized linear model framework. Extended versions that are able to account for overdispersion as the negative binomial model have been considered by 
\citet{tempelman1996mixed}, \citet{molenberghs2007extended}, \citet{zhang2017negative}.
In psychometrics also the Conway-Maxwell-Poisson model has been used albeit without covariates  \citep{ forthmann2019revisiting}.

In the mixed thresholds model the discrete density or  mass function  is given by
\begin{align*}
f_{ij}(0)&= 1-P(Y_{pi} > 0)=1-F( \zb_{ij}^T\bb_i+\xb_{ij}^T\betab_j-\delta_{j}(0))),\\
f_{ij}(r)&= P(Y_{ij} > r-1)- P(Y_{ij} > r)=\\
&=F( \zb_{ij}^T\bb_i+\xb_{ij}^T\betab_j-\delta_{j}(r-1)) )-F( \zb_{ij}^T\bb_i+\xb_{ij}^T\betab_j-\delta_{j}(r)) ),  r=1,2,\dots
\end{align*}


A thresholds function that ensures that the responses have support $0,1,\dots$ is the shifted logarithmic thresholds function 
\[
\delta_{j}(y)= \delta_{0j}+ \delta_j \log(1+y).
\]
The resulting model is as flexible as models that account for overdispersion. 
In particular the varying slopes $\delta_j$ make it very flexible and able to account for changes in distributional shape over measurements.


\subsubsection*{Epileptics Data}

The response in the data set epil from  R package  mass is the number of seizures in a fixed period (four periods considered). As covariates we use age, treatment (1: treatment, 0: placebo) and base (number of seizures at the beginning of the trial).
The model uses  a logarithmic thresholds function $\delta_j(y)=\delta_{j0}+\delta_{j}\log(1+y)$. As response function $F(.)$  we used again the normal, the Gompertz and the Gumbel  distribution. As the following table shows the Gumbel distribution shows much better fit than the normal distribution, the fit of the Gompertz was much worse (not given).  

\begin{table}[H]
\centering 
\begin{tabularsmall}{lllrrrrrrcccccccccc}
  \toprule
& Response distribution    & treatment & base &  age &log-lik  &AIC\\ 
& Gumbel &-0.534 &0.059 & 0.019 &-604.520 &1233.040\\
& NV &-0.479 & 0.048 & 0.024 &-622.706 &1269.412\\
\bottomrule
\end{tabularsmall}
\end{table}

\begin{figure}[h!]
\centering
\includegraphics[width=7cm]{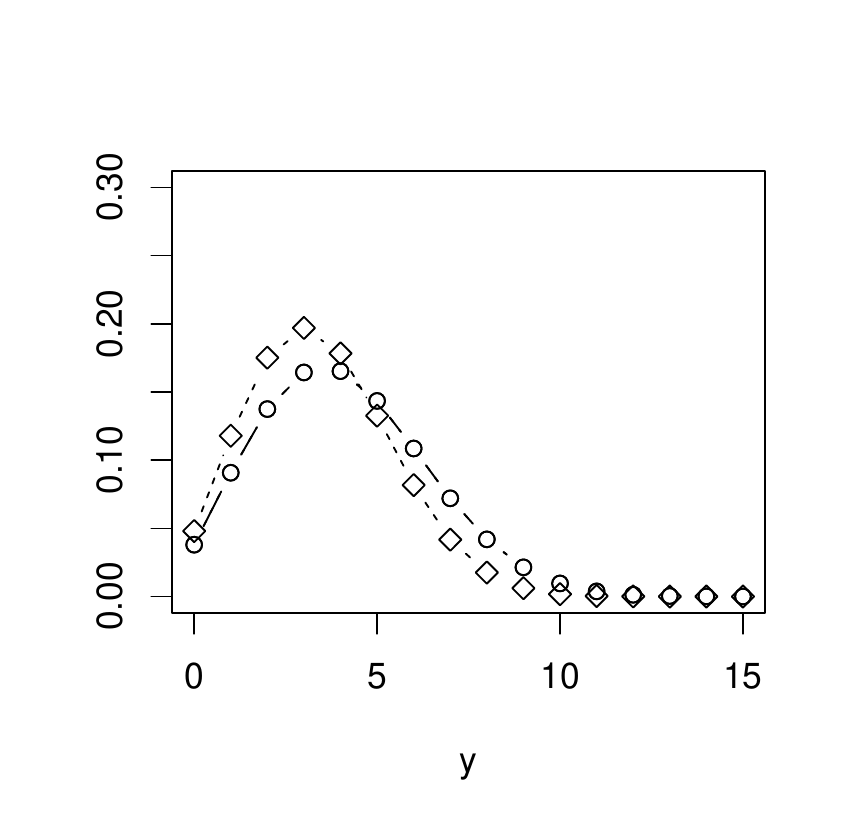}
\includegraphics[width=7cm]{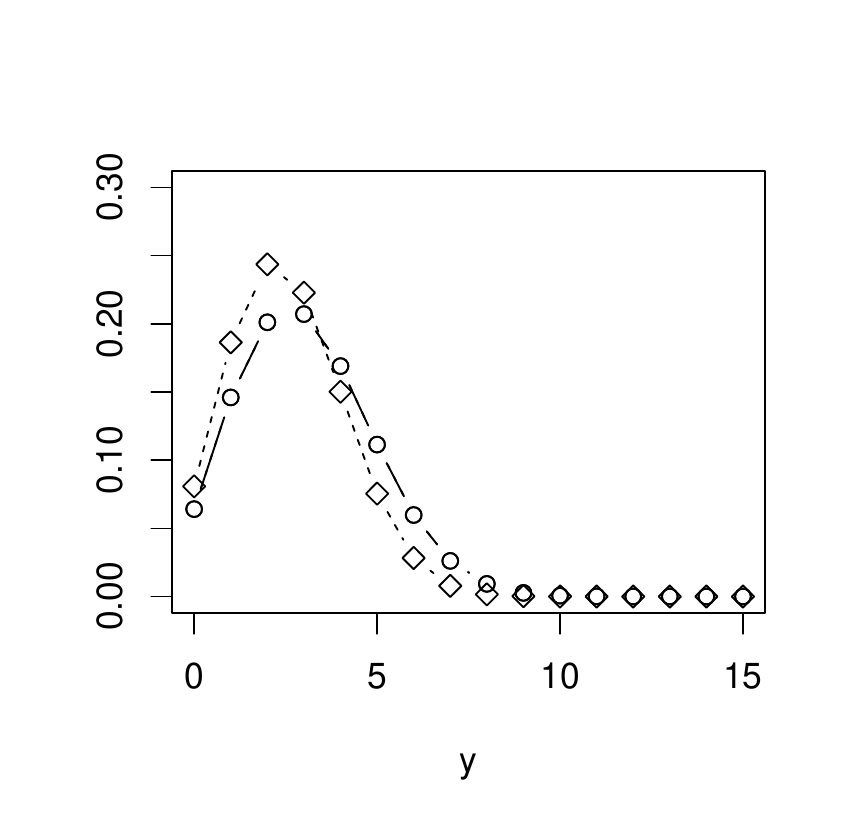}
\caption{Densities for epileptics data with $b_i=0$, base=20, age=20; left: placebo, right: traetment, diamonds indicate period four, circles period two. }
\label{fig:epp1}
\end{figure}

The  treatment effect is more pronounced in the Gumbel model yielding the likelihood-ratio test 5.658 as compared to 3.29 if the normal distribution is used. 
For illustration Figure \ref{fig:epp1} shows the densities for period 2 and 4 for placebo (left) and treatment (right) when using the Gumbel distribution, The other variables were chosen by $b_i=0$, base=20, age=20, diamonds indicate period four, circles period two. It is seen that treatment distictly reduces the number of seizures.

The thresholds model fits much better than the Poisson model, which yields log-likelihood 
-651.8071 and AIC 1313.614. It also fits better than the more general 
negative binomial model, which yielded log-likelihood -609.711.   AIC  values of the Gumbel thresholds model and the negative binomial model are comparable (AIC for negative binomial model: 1231.423. Fitting of the Poisson and the negative binomial model was done by using the R package glmmTMB \citep{brooks2017glmmtmb}.

\subsection{Random Effects Models for Ordered Responses} 

Let $Y_{ij}$ take values from $\{1, \dots,k\}$ and assume that categories are ordered. 
The typical mixed model for this type of data is the  cumulative mixed model considered among others by \citet{Jansen:90} and
\citet{TutHen:96} for univariate random effects. It has the form
\begin{align}\label{eqi:cum}
P(Y_{ij} > r|\bb_i,\zb_{ij},\xb_{ij})=F(  \beta_{0r} +\zb_{ij}^T\bb_i+\xb_{ij}^T\betab)), \quad r=1, \dots, k-1
\end{align}
with ordered intercepts $\beta_{0r} \ge \beta_{0,r+1}$ for all $r$.
For random intercepts $\zb_{ij}^T\bb_i=b_i$ it can be fitted by using the package \textit{ordinal} .

The model is equivalent to the thresholds model
\begin{align}\label{eqi:cumthr}
P(Y_{ij} > r|\bb_i,\zb_{ij},\xb_{ij})=F(\zb_{ij}^T\bb_i+\xb_{ij}^T\betab_j-\delta_{j}(r)))
\end{align}
with the special choices $\delta_{j}(r)=-\beta_{0r}$ and $\betab_1=\dots=\betab_m=\betab$.

However, the  model (\ref{eqi:cumthr}) without restrictions is more general than the simple cumulative model (\ref{eqi:cum}). 
In the simple cumulative model the  thresholds $\beta_{01}  \ge\dots\ge\beta_{0,k-1} $  do not depend on the measurement $j$. It is implicitly assumed that only covariates $\xb_{ij}$ modify the distribution of the dependent variables. This is far too restrictive in many applications. In addition, in model (\ref{eqi:cum}) the effect of covariates does not depend on the measurement ($\betab_1=\dots=\betab_m=\betab$). This is hardly realistic, in particular if dependent variables refer to different variables as in the fears data example considered in the following.


Since the range of outcomes is bounded similar thresholds functions as in the case of bounded continuous data should be used. We use the logit type function
 \[
\delta_{j}(y)= \delta_{0j}+ \delta_j \log((y-a)/(b-y)).
\]
with $a<1$ but close to 1, $b=k$.
grid or optimization, but very stable over varying values of $a$.

An advantage of using a threshold function of the form $\delta_{j}(y)= \delta_{0j}+ \delta_j g(y)$  instead of letting all intercepts vary freely (apart from order restrictions) is that sparser representations are obtained. Only $2m$ parameters are needed instead of $m(k-1)$, also order restriction problems do  not occur.

\subsubsection*{Fears Data}
As an illustrating example, we consider  data from the German Longitudinal Election Study (GLES).  The data   originate from the pre-election survey for the German federal  election  in  2017 and are concerned with political fears. The participants were asked: ``How afraid are you due to the ...'' - (1) refugee crisis?
- (2) global climate change?
- (3) international terrorism?
- (4) globalization?
- (5) use of nuclear energy?
 The answers were measured on Likert scales from 1 (not afraid at all) to 7 (very afraid). 

We fitted a discrete thresholds with logistic response function and logit difficulty function including covariates, gender (1: female; 0: male), standardized age in decades, EastWest (1: Eastern German countries/former GDR, 0: Western German countries/former FRG) and Abitur (High school degree for the admission to the university, 1:yes, 0:no). 
Table \ref{tab:fearscov1} shows the estimates of item parameters.   The parameters given are the intercept and the slope of the difficulty function  $\delta(y)=\delta_{0j}+\delta_{j}\log((y-a)/(b-y))$.
It is seen that   all items show significant covariate effects for at least one of the covariates ( z-values given below estimates).  Older respondents tend to be more afraid than younger respondents, in particular concerning terrorism but less concerning climate change. Females have for all items higher fear levels than males, the effects of EastWest are rather mixed,  people from the Eastern parts of the country are more afraid of globalization but less afraid of nuclear energy.
Higher education seems to reduce the level of fears.
The necessity of covariates is also supported by testing. The log-likelihood test that compares the model without covariates to the model with covariates is 88.081on 10 df. Thus, the covariates turn out to be influential if one accounts for the heterogeneity in the population. 

The  model fits better than the common cumulative model (\ref{eqi:cum}), in which thresholds do not depend on $j$. By default the package \textit{ordinal} fits models with global covariate effects, that is, $\betab_1=\dots=\betab_{k-1} =\betab$. By constructing appropriate design matrices it is possible to fit variable-specific covariate effects. The corresponding model has log-likelihood -1710.76 which is much smaller than -1684.871 for the thresholds model. Consequently, in terms  of AIC values the  thresholds 
model fits better. AIC value for the model with measurement-specific covariate effects but global thresholds 
effects is 3475.52 (20 covariate effects, 6 thresholds, variance of mixing distribution). For the thresholds model one obtains 3431.87 (20 covariate effects, 10 difficulty function parameters, variance of mixing distribution).

\begin{table}[h!]
 \caption{ Parameter estimates for the fears data with logit difficulty function, logistic  response function,   z-values of parameter estimates of covariate parameters are given in the lower part, variable age was standardized. } \label{tab:fearscov1}
\centering 
\begin{tabularsmall}{lllrrrrrrcccccccccc}
  \toprule
 &\multicolumn{2}{c}{ } &\multicolumn{4}{c}{ Parameters } &\multicolumn{2}{c}{ }\\
 \midrule
 &&Item  & intercepts & slopes & Age & Gender & EastWest & Abitur \\ 

\midrule
& & & & & measurement-&specific &effects\\
\midrule
&1&refugee& -0.341 &1.653        &0.151 &0.606 &0.392 &-1.388 \\
&2&climate change& -1.140 &1.858 &0.013 &1.061  &-0.594 &-0.062 \\
&3&terrorism& -2.062 &1.795      &0.393 &1.329 &0.321 &-1.207 \\                  
&4&globalization& 0.716 &1.862   &0.195 &1.020 &0.725 &-0.719\\
&5&nuclear energy& -0.922 &1.641 &0.254 &0.379  &-0.416 &-0.300 \\
\midrule
&&Log-lik &-1684.871&&\\
\midrule
& & & & &&$z-$values &covariates\\
\midrule
&1&refugee&  &       &0.937 &1.883  &1.179 &-4.002 \\
&2&climate change& & &0.079 &3.313  &-1.798 &-0.186 \\
&3&terrorism& &      &2.354 &3.935 &0.939 &-3.468 \\
&4&globalization& &  &1.236 &3.184 &2.206 &-2.108 \\ 
&5&nuclear energy& & &1.565 &1.189  &-1.253 &-0.892 \\
\midrule
& & & & & &global &effects\\
\midrule
& &Log-lik &-1715.82  & &0.230 &1.030 &0.290 &-0.545 \\
\bottomrule
\end{tabularsmall}
\end{table}

\section{ Joint Modeling of Different Types of Responses}\label{sec:joint}

A strength of the thresholds model is that dependent variables can be of various types. It allows for some of the measurements to be continous while others are binary, ordinal or given as  counts. Also the combination of continuous measurements with differing support  can be modeled in a joint random effects model. There have been some approaches to joint modeling of different types of responses in specificsettings, see, for example, \citet{ivanova2016mixed} with a focus on ordinal variables or \citet{loeys2011joint}, where a joint modeling approach for reaction time and accuracy in psycholinguistic experiments has been proposed.
However, no general random effects model that allows to combine different types of responses seems available

The flexibility of thresholds models to account for various types of measurement is due to the general form of the model. Since the same model form, which specifies $P(Y_{ij}> y)$, applies to different types of measurement it is straightforward to obtain a joint model simply by allowing for different distributions (continuous or discrete) and specifying the thresholds function accordingly. For example, if measurement 1 is continuous and measurement 2 ordered categorical, one can choose for the first measurement the linear or logarithmic thresholds function (depending on the support) and for the second measurement the logistic thresholds function. The common random effects will account for the correlation between measurements without the need for an explicit new concept for the correlation between a continuous and an ordered categorical variable.

\subsubsection*{Sleep Data}
As an illustrating example let us again consider the sleep deprivation data. Instead of using the average reaction time per day in all 10 measurements, the last two measurements were 
transformed to ordered categorical data. More concrete, the interval $ (200,500)$, which covers the reaction times, has been divided into six equidistant intervals and  responses  
were coded as $1,\dots,6$ according to the responses in intervals. For the first eight measurement the logarithmic threshold function has been chosen, and they are specified as continuous. For the last two measurements the logit thresholds function has been chosen and the measurements are specified as discrete with values $1,\dots,6$. The fitting of the model with normal response function yielded log-likelihood -768.597, which has to differ from the likelihood of the model with continuous responses considered earlier (-875.50) 
since now a combination of continuous and discrete  distributions is assumed.  

\begin{figure}[h!]
\centering
\includegraphics[width=7cm]{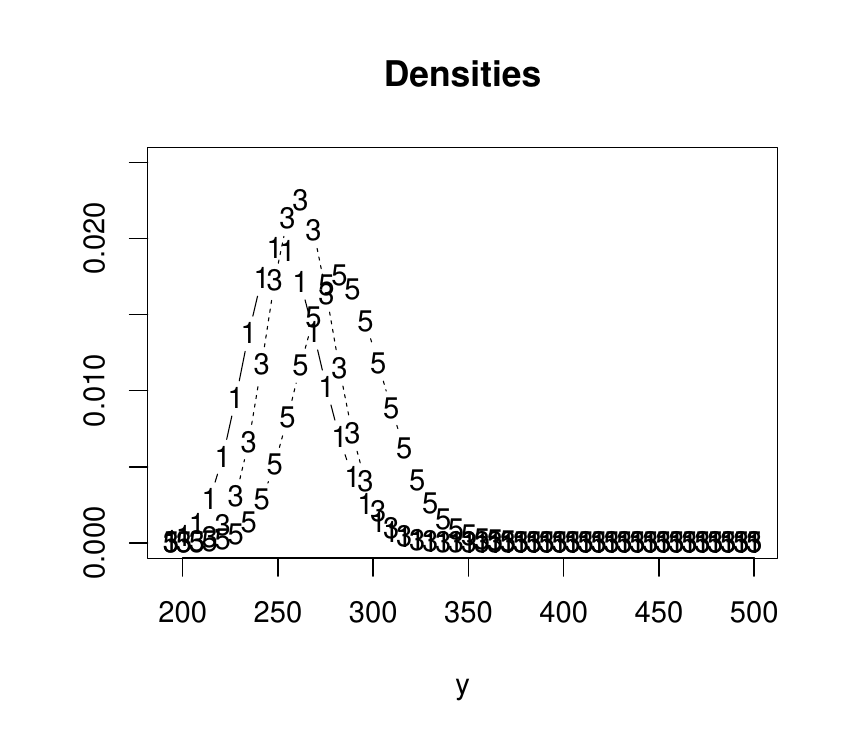}
\includegraphics[width=7cm]{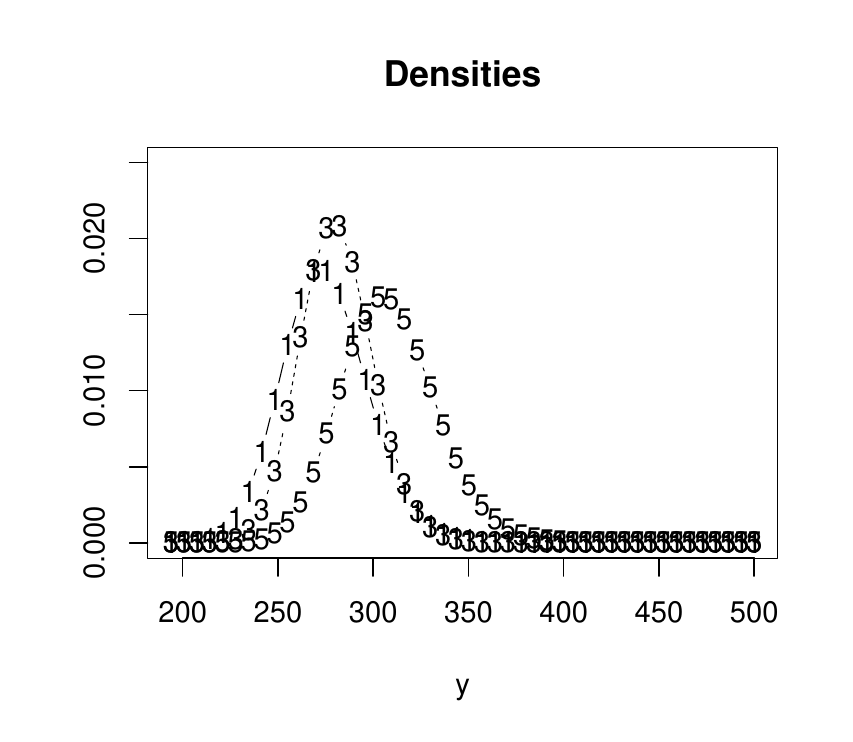}
\caption{Reaction time for days 1,3,5  of sleep deprivation for $b_i=0$ (left) and $b_i=1$ (right) for mixed responses.}
\label{fig:sleepmixed}
\end{figure}

However, the variation of reaction times over  days remains essentially the same. Figure 
\ref{fig:sleepmixed} shows the densities for days 1,3,5   for $b_i=0$ (left) and $b_i=1$ (right). They are practically the same as the densities given in Figure \ref{fig:sleep1}, which shows the densities if all variables are considered as continuous.

\section{Estimation}\label{sec:est}

A general form of the 
 threshold function term is given by 
  $\delta_{j}(y))= \Phib_j(y)^T\deltab_j$, where $\Phib_j(y)=(\Phi_{j0}(y),\dots,\Phi_{jM}(y))$ is a vector that contains functions of $y$. If the same threshold function is used for all measurements it can be specified as $\Phib_j(y)^T=(1,g(y))$. Measurement-specific thresholds functions can be obtained by $\Phib_j(y)^T=(1,g_j(y))$. But also more general vectors of functions can be useful. 
Using this parameterization, the model has the form
\begin{align}\label{eq:modest}
P(Y_{ij} > y|\bb_i,\zb_{ij},\xb_{ij})=F(\zb_{ij}^T\bb_i+\xb_{ij}^T\betab_j-\Phib_j(y)^T\deltab_j),
\end{align}

For continuous measurement $j$ the  density is
\begin{align*} 
f_{ij}(y)=f(\zb_{ij}^T\bb_i+\xb_{ij}^T\betab_j-\delta_{0j}- \Phib_j(y)^T\deltab_j)\Phib_j^{'}(y)^T\deltab_j, 
\end{align*}
where $\Phib^{'}_j(y)=(\Phi^{'}_{j0}(y),\dots,\Phi^{'}_{jM}(y))$ contains the derivatives of the components and $f(.)$ is the density linked to $F(.)$.
If $\Phib_j(y)^T=(1,g(y))$ one obtains the simpler form
\begin{align*} 
f_{ij}(y)=f(\zb_{ij}^T\bb_i+\xb_{ij}^T\betab_j-\delta_{0j}- \Phib_j(y)^T\deltab_j)\delta_j g^{'}_j (y). 
\end{align*}

For discrete data with $Y_{ij} \in \{  0,1, \dots\}$. 
 the discrete density   is given by
\begin{align*}
f_{ij}(0)&= 1-P(Y_{pi} > 0)=1-F( \zb_{ij}^T\bb_i+\xb_{ij}^T\betab_j-\Phib_j(0)^T\deltab_j),\\
f_{ij}(r)&= P(Y_{ij} > r-1)- P(Y_{ij} > r)=\\
&=F( \zb_{ij}^T\bb_i+\xb_{ij}^T\betab_j-\Phib_j(r-1)^T\deltab_j)-F( \zb_{ij}^T\bb_i+\xb_{ij}^T\betab_j-\Phib_j(r)^T\deltab_j ),  r=1,2,\dots
\end{align*}
When assuming that random effects are normally distributed with mean zero and covariance matrix $\Sigmab$ the marginal log-likelihood has the form
\begin{align*} 
L(\{  \betab_j\}, \{  \deltab_j\} )= \prod_{i=1}^n \int \prod_{j=1}^m f_{ij} (y_{ij}) f_{0,\Sigmab}(\bb_i)d\bb_i, 
\end{align*}
where $f_{0,\Sigmab}(.)$ is the density of the normal distribution with mean zero and covariance matrix $\Sigmab$.
The corresponding  log-likelihood is given by

\begin{align*} 
l(\{  \betab_j\}, \{  \deltab_j\})= \log(L(\{  \betab_j\}, \{  \deltab_j\} ))=
\sum_{i=1}^n \log(\int \prod_{j=1}^m f_{ij} (y_{ij}) f_{0,\Sigmab}(\bb_i)d\bb_i), 
\end{align*}

Maximization of the  marginal log-likelihood can be obtained by using Gauss Hermite quadrature, which has been used in generalized mixed models, see, for example, \citet{AndAit:85}, \citet{LiuPie:94}, \citet{HedGib:94}, \citet{rodriguez2008multilevel}, \citet{gueorguieva2001multivariate}. An alternatively is  the EM, which was considered for mixed models among others by  \citet{BocAit:81} \citet{AndAit:85}. Overviews on inference tools for generalized moxed models are found in \citet{jiang2007linear}, \citet{MccSea:2001}.

\section{Finding Sparser Models}\label{sec:sparse}

As is seen from the fears data the effects of covariates can vary strongly across measurements. For example, the variable EastWest has positive values for some items but negative values for other items. People living in the Eastern part of Germany tend to show stronger fears concerning globalization and refugees but  lesser fears concerning the climate crisis than people living in the Western part of Germany. Thus, the cultural imprint from living in different countries with differing experiences seems to still be present.

The  inclusion of measurement-specific fixed effects $\betab_j$ make the general model considered here very flexible. A possible downside is that the model contains many parameters. Each variable comes with as many parameters as measurements are used. However, not each variable necessarily has effects that vary across measurements. Selecting those variables that have global effects, that is effects that do not vary across measurements, would yield sparser models with an easier interpretation.
 
One strategy to identify variables with global effect is testing. Likelihood ratio test for hypotheses of the form 
\begin{align*}
H_0:\beta_{1j}=\dots=\beta_{mj} \text{ against } H_1:\beta_{sj}\ne \gamma_{lj} \text{ for at least one pair } (s,l) 
\end{align*}
can be used to test if single variables are global.  Table  \ref{tab:fearscov2} shows the results of testing for the fears data. It is seen that all of the variables should be considered as having effects that vary over measurements. 
 
\begin{table}[h!]
 \caption{Testing if covariates can considered global, one at a time } \label{tab:fearscov2}
\centering
\begin{tabularsmall}{lllrrrrrrcccccccccc}
  \toprule
 & Age & Gender & EastWest & Abitur &log-likelihood &LR test &df\\ 
  \midrule
&var &var &var &var &-1684.871 \\
&global &var &var &var  &-1695.435 & 21.128 &3\\
&var  &global &var &var &-1708.562 &47.382   &3\\
&var &var &global &var &--1704.869 & 39.996 &3\\
&var &var &var &global &-1702.910  &36.079 &3\\
\bottomrule
\end{tabularsmall}
\end{table} 
 
 \begin{figure}[h!]
\centering
\includegraphics[width=7cm]{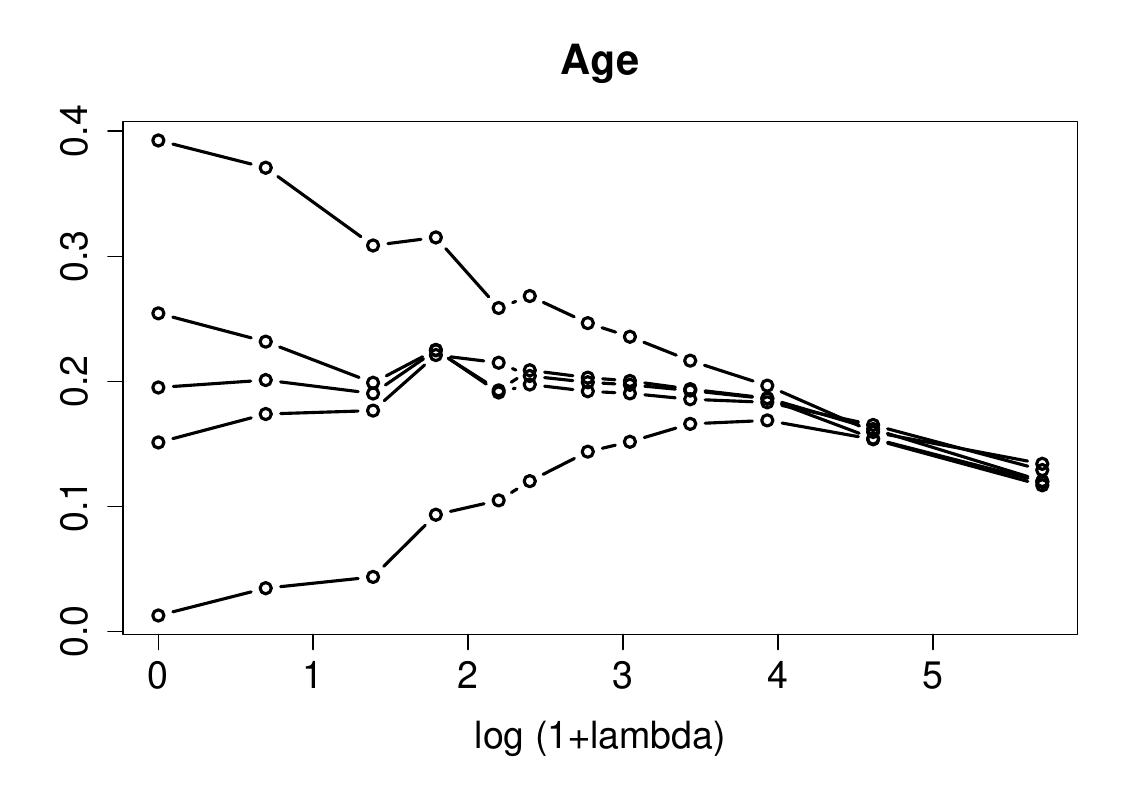}
\includegraphics[width=7cm]{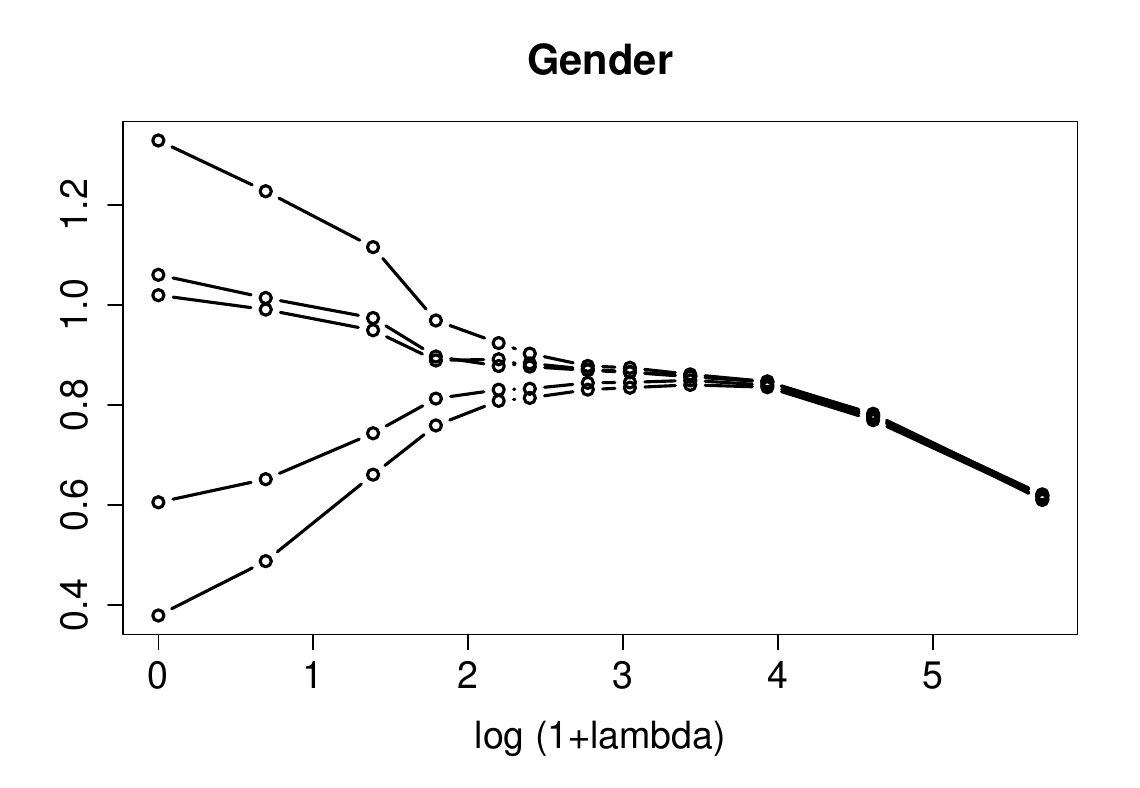}
\includegraphics[width=7cm]{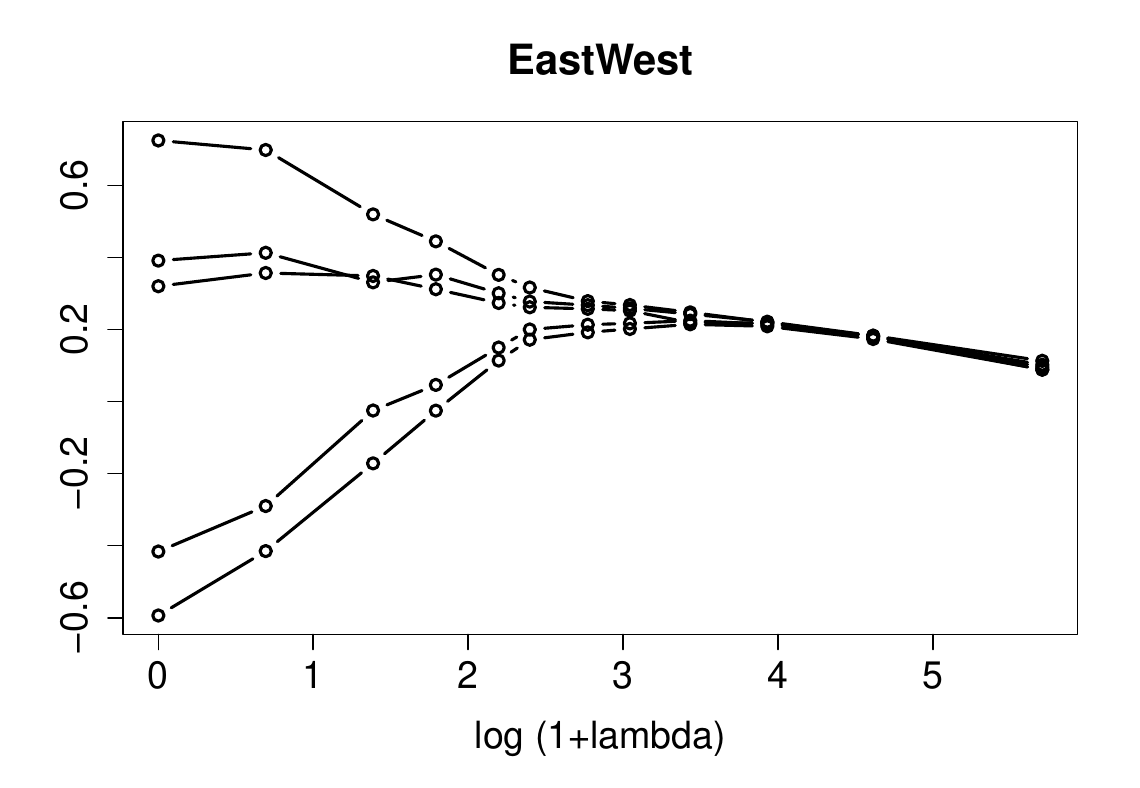}
\includegraphics[width=7cm]{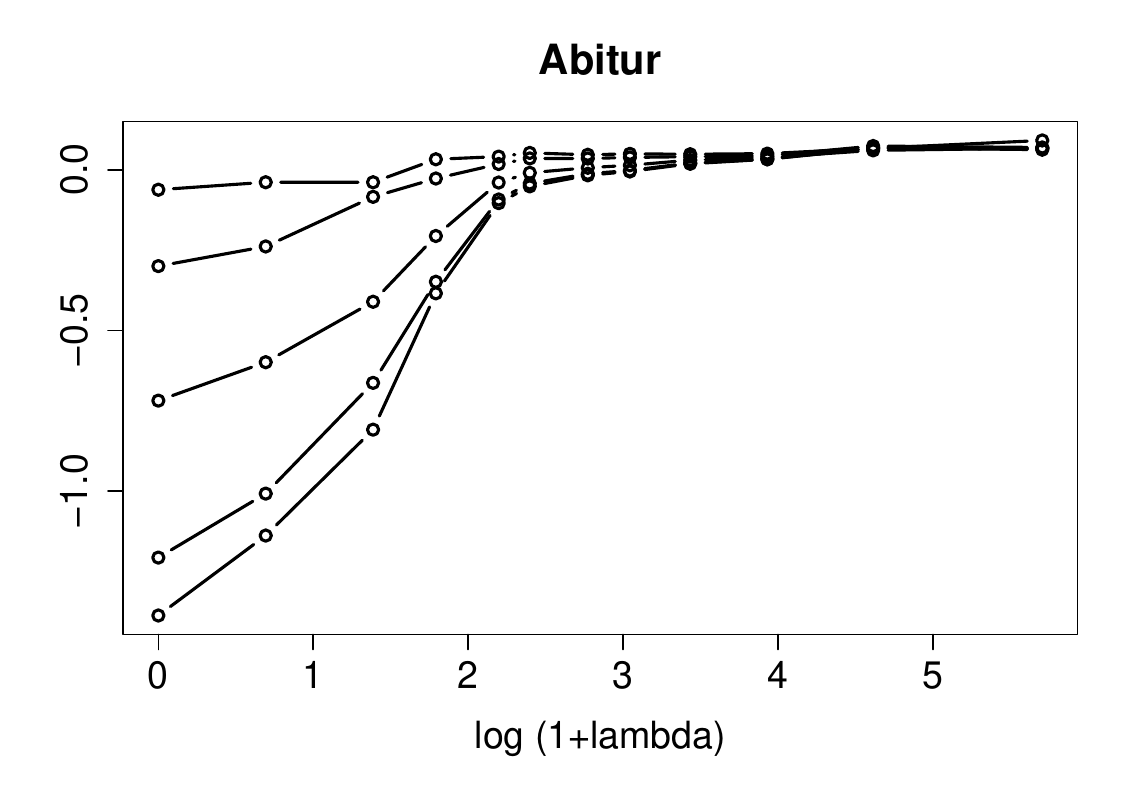}
\includegraphics[width=7cm]{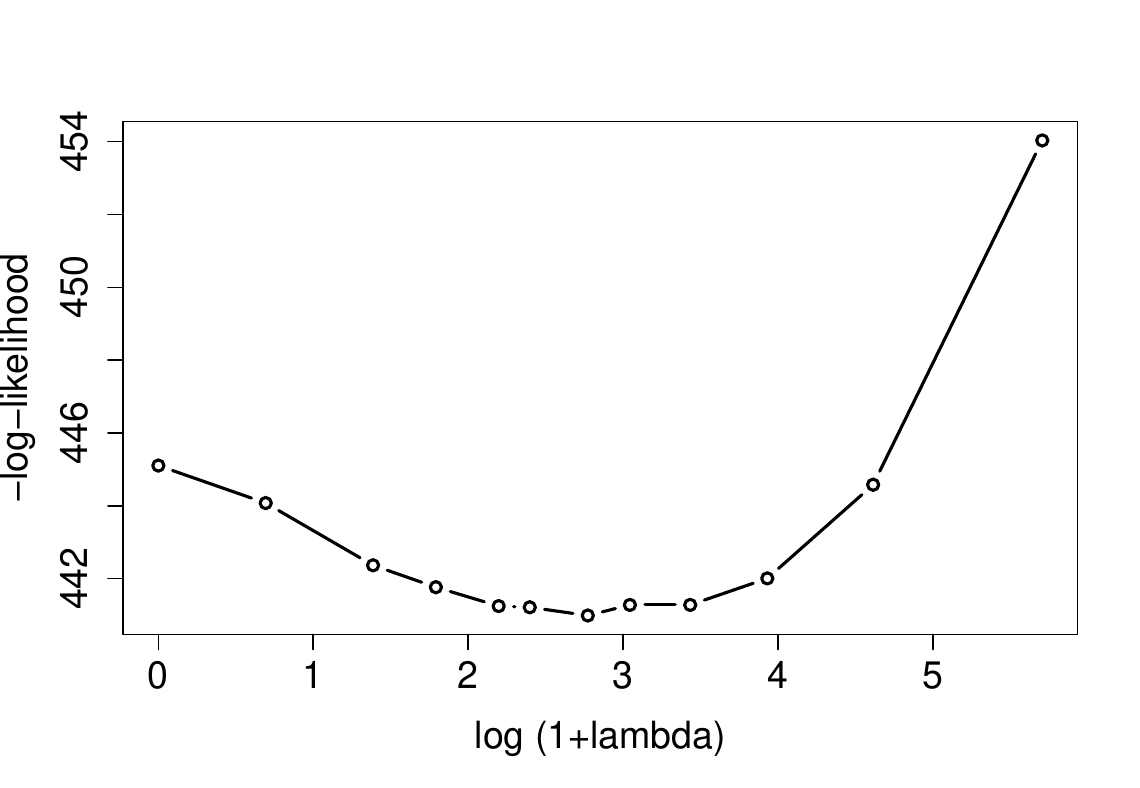}
\caption{Coefficient paths for grouped lasso on differences and selection; last panel is cross validation loss.}
\label{fig:path1}
\end{figure}
 

An alternative way to obtain sparser representations is to use penalization methods that have been widely used since the advent of the Lasso \citep{Tibshirani:96}.
Instead of maximizing the  usual marginal log-likelihood one maximizes the penalized log-likelihood
\begin{align*}
l_p(\{\betab_{j}\}, \{  \deltab_j\}) = l (\{\betab_{j}, \{  \deltab_j\})-P_{\lambda}(\{\betab_j\}),
\end{align*}
where  $l(.)$ is the usual log-likelihood and $P_{\lambdab}(\{\betab_{j}\})$ is a penalty term depending on a  tuning parameter $\lambda$.

A penalty term that enforces the building of global variables and simultaneously select variables is given by 
\begin{align*}
P_{\lambda}(\{\betab_j\})= \lambda_f (\sum_{s=1}^m  ||\betab_{s,\text{dif}}||+\sum_{s=1}^m 
||\betab_{.s}||)
\end{align*}
where $||.||$ is the $L_2$-norm, $\betab_{s,\text{dif}}^T=(\beta_{1s}-\beta_{2s},\beta_{2s}-\beta_{3s},\dots)$ contains all differences of effects between pairs of measurement for the $s$th variable, and
 $\betab_{.s}^T=(\beta_{1s}\dots,\beta_{ms})$ is the vector that collects all parameters linked to the $s$-th variable.
It is a penalty term built from two terms. The first term  enforces the building of global variables. It is  a grouped lasso type penalty that aims at the fusion of parameters. It enforces that whole groups of differences are set to zero with the group referring to differences of effects of single variables. In common grouped lasso penalties as considered by \citet{YuanLin:2006} and \citet{MeiGeeB:2007}  groups refer to parameters linked to variables variables, not differences of parameters linked to variables. Fusion penalties have been used in particular to combine effects of adjacent categories when modeling the effects of  categorical predictors, see \citet{GerTut:2016a}, \citet{gertheiss2023regularization} and to
enforce selection and shrinkage to global effects, for example, by  \citet{OelTu2017}. 
The second term is a selection term. It enforces that all parameters linked to specific variables are set to zero. 


Figure \ref{fig:path1} shows the resulting coefficient paths plotted against  $\log(1+\lambda)$. For growing $\lambda$ coefficients become more similar (global) while the shrinkage toward zero (elimination of variables) sets in rather late. Zero estimates are obtained for very large values of $\lambda$, well beyond the values that are shown. This is sensible since the variables certainly have an effect on the levels of fear. The last panel shows the negative cross-validation log-likelihood (5-fold cross-validation). It is seen that the smallest levels are obtained if  $\log(1+\lambda)$ is slightly above 2. In this range coefficients are still measurement-specific, which supports the testing results.

\section{Concluding Remarks}
It has been demonstrated that the thresholds model is very versatile and can be adapted to quite different distribution functions. Moreover, it can be used in joint modeling of different responses, which has not been considered in the literature since appropriate models have not been available. A rare exception is the modeling of survival data where joint modeling of survival times and  
longitudinal data  has been considered, see, for example, \citet{hsieh2006joint}. 

We used marginal estimation based on integration methods, but alternative estimation from the toolbox of mixed models could be used. Also alternative regularization methods could be useful when trying to find sparser representations, for an  overview on basic regularization method including boosting, see \citet{HasTibFri:2009B}.

The class of mixed thresholds models could be made even more flexible by letting the data decide which thresholds function is appropriate.  In particular, B-splines as considered extensively by \citet{eilers2021practical} could be useful, which has been  demonstrated by 
\citet{TuItThr2022} in the item-response setting without covariates.

\section*{Appendix}

For simplicity, let the model be given in the form 
\begin{align}\label{eq:di1f}
P(Y_{ij} > y|\bb_i,\zb_{ij},\xb_{ij})=F(\tilde\eta_{ij}-\delta_{j}(y)),
\end{align}
where $\tilde\eta_{ij}=\zb_{ij}^T\bb_i+\xb_{ij}^T\betab_j$ and $F(.)$ is a strictly increasing distribution function.

\begin{theorem} \label{t1} 
Let the threshold function have the form  $\delta_{j}(y)= \delta_{0j}+ \delta_j y$, $\delta_j \ge 0$. Then,   one obtains for the expectation and the variance
\begin{align*}  
&\E(Y_{ij})=\mu_{ij}=  \frac{1}{\delta_{j}}(\beta_{0j}+\zb_{ij}^T\bb_i+\xb_{ij}^T\betab_j), \\
&\var(Y_{ij})=\sigma_{ij}^2=  \frac{\sigma_F^2}{\delta_j^2},
\end{align*} 
where $\beta_{0j}=-\mu_F-\delta_{0j}$ with  $\mu_F=\int y f(y)dy$,   $\sigma_F^2=\var_F =\int (y-\mu_F)^2f(y)d y$, $f(y)=\frac{\partial F(y)}{\partial y}$.


 
\end{theorem}

Proof:
For linear item function the thresholds model has the form $P(Y_{ij} > y|\bb_i,\zb_{ij},\xb_{ij})=F( \tilde\eta_{ij}-\delta_{j}(y))$. The corresponding distribution function is
\[
F_{Y_{ij}}(y) = P(Y_{ij} \le y) = 1 - F(\tilde\eta_{ij}-\delta_{0j}- \delta_j y),
\]
which has the  density 
\[
f_{Y_{ij}}(y)=\frac{\partial F_{Y_{ij}}(y)}{\partial y} =  f(\tilde\eta_{ij}-\delta_{0j}- \delta_j y)\delta_j.
\]
Thus, the mean is given by

\[                  
\E(Y_{ij})=  \delta_j \int y f(\tilde\eta_{ij}-\delta_{0j}- \delta_j y) dy.
\]
With $\eta = \tilde\eta_{ij}-\delta_{0j}- \delta_j y$   and $d\eta/dy=- \delta_j$ one obtains                   
\begin{align*}                  
\E(Y_{pi})&=   -\frac{1}{\delta_j}\int_{\infty}^{-\infty} (\tilde\eta_{ij}-\eta- \delta_{0j}) f(\eta) d\eta
 =  \frac{1}{\delta_j}\int_{-\infty}^{\infty} (\tilde\eta_{ij}-\eta- \delta_{0j}) f(\eta) d\eta\\
& = \frac{1}{\delta_j}(\tilde\eta_{ij}-\mu_F- \delta_{0j}),
\end{align*} 
where $\mu_F=\int y f(y)dy$ is a parameter that depends on $F$  only.

The variance is given by
\begin{align*}                  
\var(Y_{pi})&=   \int (y-\frac{\tilde\eta_{ij}-\mu_F- \delta_{0j}}{\delta_j})^2 f(\tilde\eta_{ij}-\delta_{0j}- \delta_j y) \delta_j dy
= \int (\frac{\eta-\mu_F}{\delta_j})^2 f(\eta) d\eta \\
&= \var_F /{\delta_j}^2,
\end{align*}
where $\var_F=\int ({\eta-\mu})^2 f(\eta) d\eta$.


\medskip
\begin{theorem} \label{equ} \label{t2}
Let the threshold function have the form  $\delta_{j}(y)= \delta_{0j}+ \delta_jg(y)$, $\delta_j \ge 0$. Then,   one obtains 
\begin{align*} 
&\E(g(Y_{ij}))=  \frac{1}{\delta_{j}}(\beta_{0j}+\zb_{ij}^T\bb_i+\xb_{ij}^T\betab_j), 
\quad\var(g(Y_{ij}))=  \frac{\sigma_F^2}{\delta_j^2},
\end{align*}
\end{theorem}

Proof:
The mean is given by 
\begin{align*}
\E(g(y)) = \int g(y) f(\tilde\eta_{ij}-\delta_{0j}- \delta_j g(y))\delta_j g^{' }(y)dy 
\end{align*}
With $\eta= \tilde\eta_{ij}-\delta_{0j}- \delta_j g(y) $ one obtains
\begin{align*}
\E(g(y)) = -\int  \frac{\eta-\tilde\eta_{ij}+\delta_{0j}}{\delta_j}  f(\eta)  d\eta 
=\frac{\tilde\eta_{ij}-\delta_{0j}-\mu_F}{\delta_j}=\frac{\tilde\eta_{ij}+\beta_{0j}}{\delta_j},
\end{align*}
since $\beta_{0j}=\delta_{0j}-\mu_F$.

The variance is given by 
\begin{align*}
\var(g(y)) &= \int (g(y)-\frac{\tilde\eta_{ij}-\delta_{0j}-\mu_F}{\delta_j}))^2 f(\tilde\eta_{ij}-\delta_{0j}- \delta_j g(y))\delta_j g^{' }(y)dy \\
&=\int (\frac{\eta-\mu_F}{\delta_j})^2 f(\eta) d\eta 
= \var_F /{\delta_j}^2.
\end{align*}

\bibliography{literaturgeneral}  

\end{document}